\documentclass[sigplan, table, shortauthors, nonacm,11pt]{acmart}
\startPage{1}
\pdfoutput=1

\usepackage{listings}
\definecolor{codegreen}{rgb}{0,0.6,0}
\definecolor{codegray}{rgb}{0.5,0.5,0.5}
\definecolor{codepurple}{rgb}{0.58,0,0.82}
\definecolor{backcolour}{rgb}{0.95,0.95,0.92}
\title{Sionnx: Automatic Unit Test Generator for ONNX Conformance}        

\author{
  Xinli Cai,  Peng Zhou, Shuhan Ding, Guoyang Chen, Weifeng Zhang}
\affiliation{
  \institution{Alibaba Group US Inc.}            %% \institution is required
}
\email{{xinli.cai, peng.z, s.ding, g.chen, weifeng.z}@alibaba-inc.com}

\begin{document}

\begin{abstract}
Open Neural Network Exchange (ONNX) is an open format to represent AI models and is supported by many machine learning frameworks. 
While ONNX defines unified and portable computation operators across various frameworks, the 
conformance tests for those operators are insufficient, which makes it difficult to verify if
an operator's behavior in an ONNX backend implementation complies with the ONNX standard.
In this paper, we present the first automatic unit test
generator named \textit{Sionnx} for verifying the compliance of ONNX implementation.  First, we propose a compact yet complete set of rules 
to describe the operator's attributes and the properties of its operands.
Second, we design an Operator Specification Language (OSL) to provide a high-level description for the operator's syntax.
Finally, through this easy-to-use specification language, we are able to build a full testing specification which leverages LLVM TableGen to
automatically generate unit tests for ONNX operators with much large coverage.
Sionnx is lightweight and flexible to support cross-framework verification.
%By taking advantage of each framework, Sionnx provides more flexibility for the debugging process. 
The Sionnx framework is open-sourced in the github repository (https://github.com/alibaba/Sionnx).

\end{abstract}

%\ccsdesc[500]{Software and its engineering~General programming languages}
%\ccsdesc[300]{Social and professional topics~History of programming languages}
%% End of generated code

%% Keywords
%% comma separated list
%\keywords{DNN, FPGA, Compiler, Optimization, Memory}  %% \keywords are mandatory in final camera-ready submission
%\settopmatter{printfolios=true}
\maketitle

\section{Introduction}
\label{sec:intro}

\begin{figure}
  \centering
  \includegraphics[width=.33\textwidth]{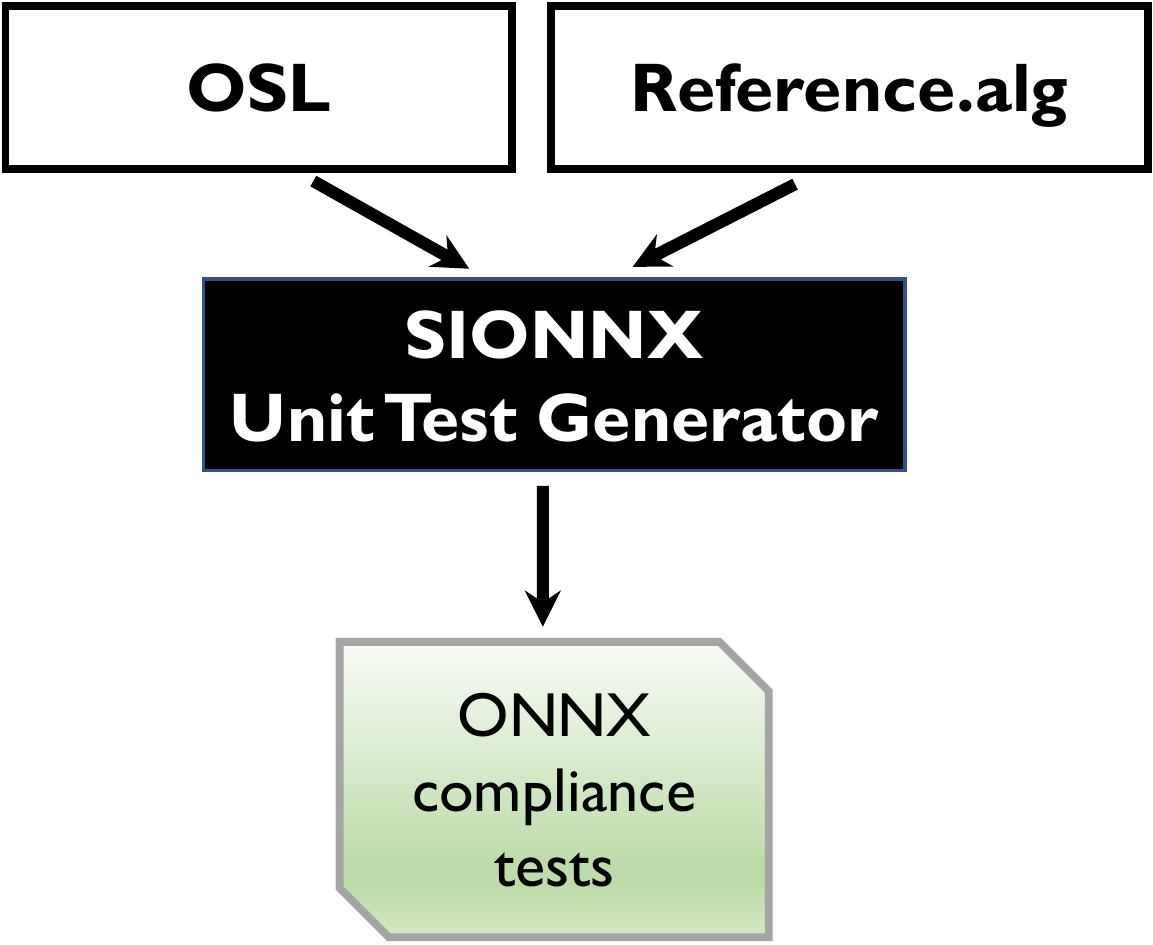}
  \caption{Overview of Sionnx}\label{fig:overview}
\end{figure}

In the past few years, Deep Neural Network (DNN) has shown its extraordinary ability in solving many complex machine learning problems, such as image
classification~\cite{Ciregan+:CVPR12,Krizhevsky+:NIPS12,Simonyan+:ICLR15,
He+:CVPR16, Russakovsky+:IJCV15}, 
speech recognition~\cite{Hinton+:SPM12,Seltzer+:ICASSP13,Dahl+:ASLP12,Deng+:ICASSP13, Yu+:ICASSP13, Li+:SLT12} and bioinformatics~\cite{Saeys+:Bioinformatics07, Min+:Bioinformatics17}. 
A machine learning model consists of a list of compute operators where the output(s) of one operator could be input operands of other operators.
To ease the burden for machine learning developers to construct, train, and conduct inference on models, 
many frameworks/tools (e.g., Tensorflow~\cite{Abadi+:OSDI16}, Pytorch~\cite{Paszke+:NIPS17}, Mxnet~\cite{Chen+:MXNET}, Caffe~\cite{Jia+:Caffe}) were proposed in recent years. While these frameworks provide convenient and efficient interfaces to build machine learning models, 
a model is often difficult to port from one framework to another due to inconsistent definitions of compute operators across frameworks.
Open Neural Network Exhange (ONNX)~\cite{ONNX}, as one of the solutions to address the model portability problem, is an open format to construct machine learning models and  supported by many machine learning frameworks. Thus, models in the ONNX format can run on any of the supporting frameworks seamlessly.

In ONNX, a well-defined set of operators in machine learning are selected to represent the common behaviors of computation in the same category (e.g., convolution, pooling and matrix multiplication) across different frameworks. However, in the current ONNX framework, conformance testing for each operator relies only on a small number of handwriting test cases. This is certainly insufficient for the correctness verification and hence makes it difficult to test a framework in question against the ONNX standards for each operator's behavior.

There have been some works on how to generate unit tests automatically for extending conformance test sets in other domains. These works mainly focus on automatic unit test generation for programming languages~\cite{EvoSuite, Symstra, Testful, Randoop, Pex}. However, none of them can be directly applied here due to the major challenges as follows. 

The first challenge is how to systematically generate valid tests for different operators while satisfying their own constraints. For tools to automatically generate valid tests, the key characteristics to identify are the attributes of both the operator's operands and the operator itself. However, the constraints for those attributes can be totally different from one operator to another. For example, operator \textit{Asin}, which calculates the arcsine (inverse of sine) of the given input, has the constraint that the input value must be in the range of [-1, 1]. However, operator \textit{split}, which slices an input tensor into a list of tensors along the specified `axis', doesn't have the constraint on the value range of the input but requires the value of `axis' not exceed the number of dimensions of the input.

The second challenge is how to produce the reference result for each test. In the ONNX conformance testing system, a reference result is generated by calling a single or a list of numpy~\cite{numpy} operators. But using only numpy operators to write a reference algorithm for some ONNX operators is tedious and error-prone. Take operator LSTM as an example, to emulate the computation of a LSTM cell, more than 60 lines of Python code with Numpy are written by the developers of the ONNX framework~\cite{ONNX}. With more and more operators being added to ONNX due to its feature expansion, the burden of writing complex reference algorithms is becoming much heavier. Therefore, there is an increasing demand for finding a more practical solution in the ONNX community.

The third challenge is how to design an efficient randomization strategy to generate a limited number of tests but with large test coverage. In some cases, randomly generating more tests without any guidance doesn't necessarily improve the coverage~\cite{coverage}. In fact, there are some important points that are more likely to yield diverse behaviors of the code. For instance, for binary operators, the length of data dimensions equaling 1 is the key to trigger the broadcasting behavior. How to identify those points and find a general way to cover them in a limited number of tests are challenging.

In this paper, we propose the first automatic unit test generator for ONNX operators. Figure~\ref{fig:overview} shows the high level overview of Sionnx workflow. It contains three key components: a specification language to describe the key characteristics of a operator; a reference algorithm to produce the reference results for the correctness verification; a unit test generator which takes OSL and the reference algorithm as inputs and automatically generates the conformance tests with large test coverage through our newly designed three-phase randomization algorithm called TDBc-gen. More details will be presented in the following sections.

Overall, this paper makes the following contributions:
\begin{itemize}
    \item It proposes a compact set of rules which are concise to describe the attributes and constraints of operators. 
    \item A novel specification language named OSL is designed to fully describe the characteristics of the operators in a systematic way. 
    \item It can leverage some well-established frameworks to help verify the correctness of operators.
    \item It demonstrates that Sionnx is extremely lightweight and able to generate test cases for ONNX operators with large test coverage.
\end{itemize}

%\section{Overview of Sionnx}

\section{Operator Specification Language}
\begin{figure*}[h]
  \centering
  \includegraphics[width=1.\textwidth]{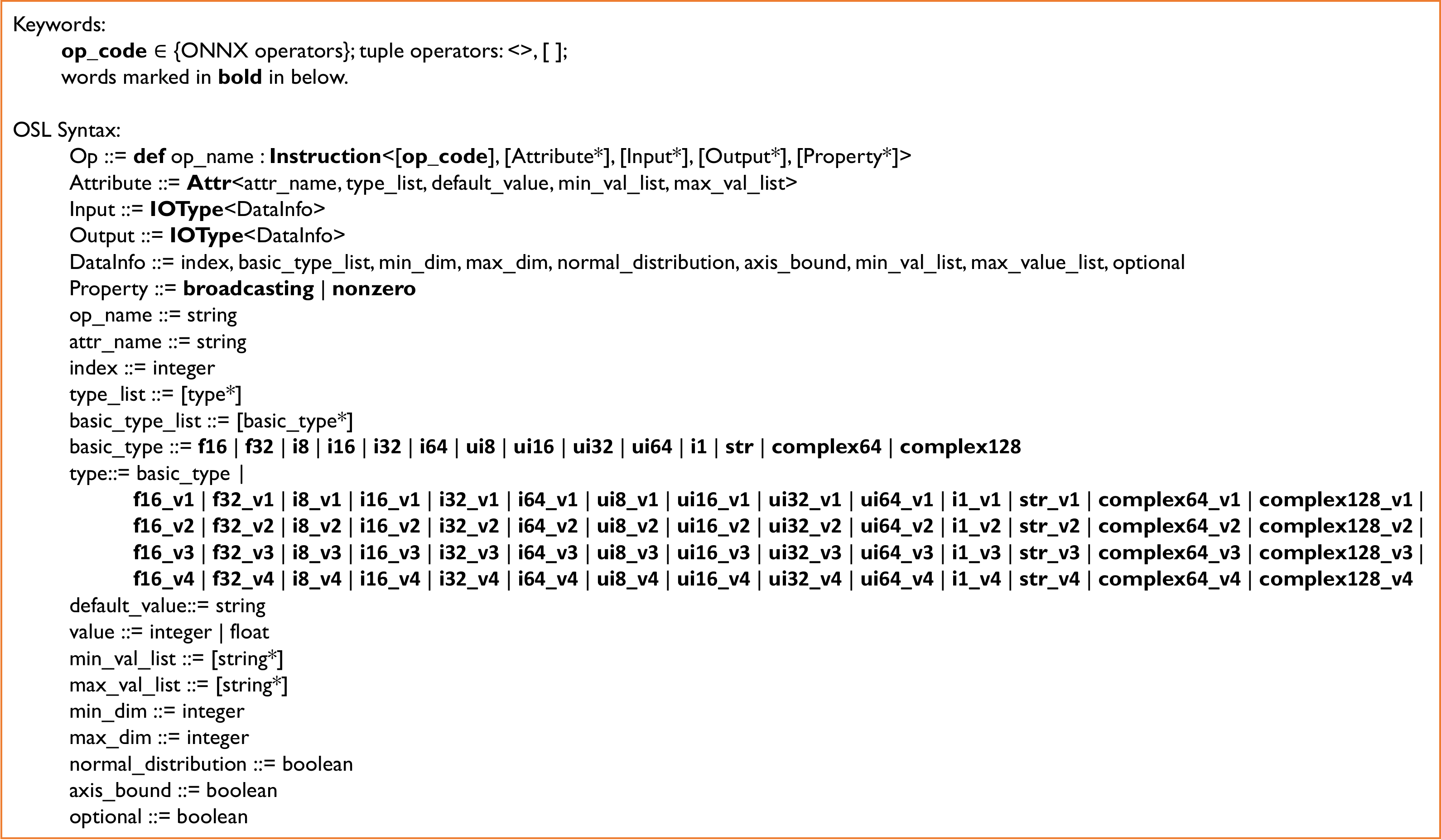}
  \caption{Syntax of OSL}\label{fig:osl-syntax}
\end{figure*}

\begin{figure*}[h]
  \centering
  \includegraphics[width=1.0\textwidth]{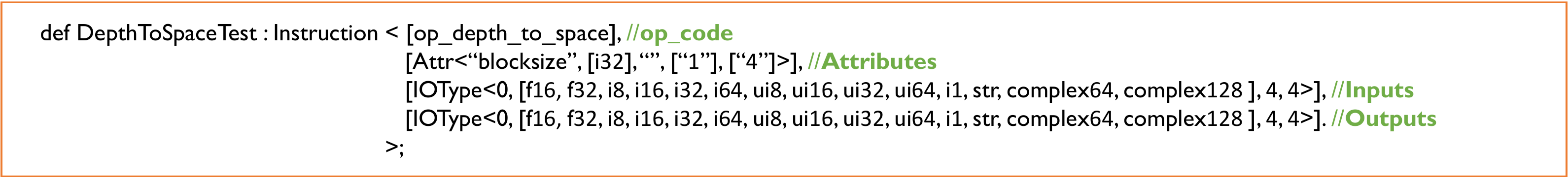}
  \caption{The operator specification of DepthToSpace in OSL}\label{fig:osl-example}
\end{figure*}

OSL is a small description language to describe the operator's characteristics including the constraints such as the requirements of operands' types, shapes and value ranges. It is worth noting that those specifications are sufficient for Sionnx generator to exploit to generate valid operands and configurations of the operator.

\subsection{Core Rules for Data Properties: An Insight}
%We conduct an empirical study on the ONNX operators and conclude a core set of specifications to represent the operator. 
The major challenge of designing OSL is how to allow a simple yet structured expression to describe those various characteristics. Our solution is based on the insight that those characteristics are essentially about data properties in three main aspects: data type, data dimension and data value range. For those two examples discussed in Section~\ref{sec:intro}, the constraint that the operands of operator \textit{arcsin} must have values in the range of [-1, 1] could be expressed by the rule of data value range directly. Similarly, the constraint of operator \textit{split} that the value of attribute `axis' should not exceed the number of dimensions of the operand could be expressed by two rules as follows. 
\begin{itemize}
    \item One indicates the constraints for attribute `axis': 1) the data type is an integer; 2) the data dimension is 0, which means the attribute is a scalar; 3) the data value range (e.g., [0,3]);
    \item The other states the dimension of input has a lower bound `axis' + 1.
\end{itemize}  

We conduct an empirical study on all ONNX operators. All of their constraints can be expressed by the three data properties, which confirms the effectiveness of our core rules.

\subsection{OSL Syntax}
With the simple yet effective core rules for data properties being determined, the next step is to formalize them into a specification language. 

To exploit existing tools for designing a new domain specific language (DSL), we follow the syntax of LLVM TableGen~\cite{tablegen} for defining the specifications of OSL and customize the LLVM TableGen backend to support the newly designed language.
%but with modifications to add definitions for new keywords. 
TableGen~\cite{tablegen} is a popular tool written in C++ from LLVM framework~\cite{llvm}. It provides interfaces to define flexible description for the specification and makes it easier to structure domain specific information.

Figure~\ref{fig:osl-syntax} shows the syntax of the OSL with some token rules such as integer, string and basic data types(e.g., f16, f32, f32\_v1) omitted. At a high-level view, an OSL specification contains one entry for test name, one entry for operator code, a list of entries for attributes, inputs, outputs and properties. The field op\_name specifies the prefix name of tests to be generated. The field op\_code indicates the operator code. The candidate is from the ONNX operator set~\cite{OnnxOP}.

\paragraph{Attribute Specification} The attribute specification corresponds to the attributes of the operator itself. For example, the padding mode for convolution~\cite{conv}, the axis for concatenation~\cite{Concat} and so on. It contains five fields. The field `attr\_name' gives the attribute name, which is a string. The string must be exactly the same as the attribute name of the operator in the ONNX standard. The next field is `type\_list'. Only types in this list are allowed as the data type of the attribute. The field `default\_value' specifies the default value of the attribute. Note that if a string is specified, the real value will be calculated according to the data type of the attribute during the interpret time. Thus, we can just use one value for different data types for simplicity. The fields `min\_val\_list' and `max\_val\_list' are designed to cooperatively indicate a list of discrete data value ranges. Elements of same index in the two lists form a pair \{min\_val\_list[index], max\_val\_list[index]\}, indicating a valid data range. It is required that the lengths of `min\_val\_list' and `max\_val\_list' must be equal. For example, if min\_val\_list = [`20',`50',`90'] and max\_val\_list = [`30',`60',`120'], then the valid data value can only be in range [20, 30] or [50, 60] or [90, 120].

\paragraph{Input and Output Specifications} In our design, the definitions of input and output specifications are same since both the inputs (or operands in the instruction terminology) and the outputs of a operator are essentially tensor data. Furthermore, an output of a operator could be an input of another operator. Hence, there are no  differences in describing the data properties for both. However, in order to distinguish which specifications are for the inputs and which are for the outputs, our design requires a list of input specifications followed by a list of output specifications. In an input/output record, 9 fields need to be filled. To illustrate the meanings of them, we select four most complex fields with detailed explanation. 

\begin{itemize}
    \item \textbf{index}\\
    It indicates that the specification is for the \$index'th input/output. The value is normally an integer no less than 0. However, if the value is -1, it means there may be multiple inputs/outputs that need to be concatenated as one single input/output.\\
    
    \item \textbf{basic\_type\_list}\\
    It indicates the legal data element types (e.g., f16, f32, i8, ...) for the tensor data. The difference between basic\_type\_list here and type\_list in the attribute specification is that the latter indicates not only the data type but also the data shape.\\
    
    \item \textbf{axis\_bound} \\
    If it is set true and there is an attribute named `axis', then the number of dimensions of the tensor data should be no less than the value of `axis' + 1. This field ensures the correct dependency between the number of dimensions of tensor data and the attribute's value. \\
    
    \item \textbf{optional} \\
    This field indicates that the tensor data may not be used with a 1-in-2 chance. 
    
\end{itemize}
For other fields, the fields min\_dim and max\_dim indicate the range of the number of dimensions; the field normal\_distribution, if set true, indicates that the tensor data will be in a normal distribution with the mean being 0 and the standard deviation being 1.

\paragraph{Other Implicit Properties of the Operator} In addition to the explicit properties (e.g., padding mode, convolution strides) of the operator that we need to describe in the attribute specification, there are some implicit properties and they can be expressed in the property specification in OSL. Different from the properties which need to be set explicitly (otherwise the default value will be used), the implicit properties are the built-in features and come into effect whenever their conditions are met. Currently, there are two implicit properties identified for ONNX operators in OSL: broadcasting and nonzero. Take operator \textit{Add} as an example, it computes the element-wise addition of two operands. When the operands have the same number of dimensions and the length of each dimension is either a common length or 1., the implicit property of broadcasting may take effect to complete the computation depending on the broadcasting type. There are two types of broadcasting. One is called `Multidirectional Broadcasting' and it is supported by 15 operators including Add, And, Div and so on. The other one is `Unidirectional Broadcasting' but only supported by two operators: Gemm and PRelu. More details about broadcasting can be found in  reference~\cite{broadcast}. %\TODO{The other implicit property is `nonzero'. what is nonzero??}

\paragraph{Corner Cases} There are some corner cases for  operators having complex dependencies among their operands and attributes. For instance, in operator "DepthToSpace", the length of dimension along the channel/depth axis of the input data must be a multiple of the value of attribute "blocksize". Such constraint is often particular for that operator and expressing it in OSL will make the design of OSL much more complex. With careful consideration of the simplicity and the expressibility in OSL, we determine to keep the simple form of OSL but handle those corner cases inside the compiler. The operators having such exclusive constraints are "DepthToSpace", "BatchNorm", "Compress", "Concat", "Gemm", "MatMul", "Conv", "OneHot", "Squeeze" and "LRN". The details of them can be found in the open-sourced github repository.

\begin{table*} 
\begin{lstlisting}[caption=\textbf{DepthToSpace.algorithm with Numpy} (The typical method to define a reference function in ONNX; It contains three numpy operations and their function arguments are carefully computed), label={lst:depth-to-space-numpy}]
#x_0: 0'th input; blocksize: blocksize in DepthToSpace
def DepthToSpace_compute(x_0, blocksize):
    b, c, h, w = x_0.shape
    tmp = numpy.reshape(x_0, [b, blocksize, blocksize, c // (blocksize**2), h, w])
    tmp = numpy.transpose(tmp, [0, 3, 4, 1, 5, 2])
    return numpy.reshape(tmp, [b, c // (blocksize**2), h * blocksize, w * blocksize])
\end{lstlisting}

\begin{lstlisting}[caption=\textbf{DepthToSpace.algorithm with Tensorflow} (Our proposed method to define a reference function; It leverages the well-established operator in Tensorflow framework. Only one easy-to-configure operator is used), label={lst:depth-to-space-tf}]
#x_0: 0'th input; blocksize: blocksize in DepthToSpace
def DepthToSpace_compute(x_0, blocksize):
    x_tensor = tensorflow.convert_to_tensor(x_0)
    res = tensorflow.nn.depth_to_space(x_tensor, blocksize, data_format='NCHW')
    return tensorflow.Session().run(res)
\end{lstlisting}
\end{table*}

\paragraph{An Example in OSL} To help better understand how OSL offers a simple yet effective way to express the ONNX operator, we show OSL specification of operator DepthToSpace in Figure~\ref{fig:osl-example} as an example. The operator is to rearrange the data from depth into blocks of spatial data. At line 1, it defines the test name as "DepthToSpaceTest" and indicates the operator code is "op\_depth\_to\_space". At line 2, it specifies there is only one attribute for the operator and the attribute's name is "blocksize". The type of the attribute can only be a 32-bits integer. No default value is given. The value range is [1, 4]. The line 3 indicates that there is only one input; the data type can be any of 16-bits and 32-bits float, 8-bits, 16-bits, 32-bits, 64-bits signed/unsigned integer, boolean, string, 64-bits and 128-bits complex number. The range of number of dimensions is [4, 4], indicating that the number of dimensions of the input must be 4. The reason for this strict constraint is that operator "DepthToSpace" requires the input data format as "NCHW"(N: batch axis; C: channels; H: height; W: width). It should be noted that the other fields not filled in the specification mean that there are no constraints for them. At line 4, the output specification is same as the input, which is true since the operator doesn't change the number of dimensions, nor the data element value.

\section{Reference Algorithm}
So far, we have showed how to describe the operator for Sionnx compiler to automatically generate valid inputs for operators. As mentioned in Section~\ref{sec:intro}, for the correctness verification, an algorithm to produce reference results should also be provided.

In Sionnx, we provide an interface for programmers to define a function as the reference algorithm in a .algorithm file for each operator. Currently, we only support algorithms written in Python~\cite{Python} (other language support left as future work) and make some assumptions for the definitions to support the automation of Sionnx. In the .algorithm file, the reference function to be called to generate the reference outputs should be named as "\$\{op\_name\}\_Compute". Otherwise, it will be very difficult for the test generator to tell. To be consistent, we assume the arguments for the operands are followed by the arguments for the attributes. For arguments within the same category, their orders follow either the index order of the input or the appearance order in the attribute specification.

Listing~\ref{lst:depth-to-space-numpy} gives an example of DepthToSpace.algorithm with Numpy, which is a typical way in ONNX to define the reference algorithm. x\_0 represents the 0'th input and blocksize represents the 0'th attribute. To compute the output, it does three numpy operations: two reshapes and one transpose, with careful computations for the parameters. This method requires human effort in designing the algorithm and may have poor performance due to the use of numpy without hardware accelerations.

\paragraph{Our Solution for Improvement} we propose to take advantages of well-established operators in other frameworks if possible for generating the reference outputs. Listing~\ref{lst:depth-to-space-tf} shows the version of DepthToSpace.algorithm written with Tensorflow. Instead of doing three numpy operations, it directly calls the handy DepthToSpace operator in Tensorflow, reducing human effort in the algorithm design. The performance could be also improved if hardware accelerators such as GPUs are available.

\begin{table*}
\begin{lstlisting}[caption=Automatically Generated Tests for DepthToSpace, label={lst:depth-to-space-tests}]
import numpy as np
import math
import onnx
from ..base import Base
from . import expect

class DepthToSpace(Base):

    @staticmethod
    def export():

        def DepthToSpace_compute(x_0, blocksize):
            b, c, h, w = x_0.shape
            tmp = np.reshape(x_0, [b, blocksize, blocksize, c // (blocksize**2), h, w])
            tmp = np.transpose(tmp, [0, 3, 4, 1, 5, 2])
            return np.reshape(tmp, [b, c // (blocksize**2), h * blocksize, w * blocksize])

        node = onnx.helper.make_node(
            'DepthToSpace',
            inputs=['x_0'],
            outputs=['y_0'],
            blocksize=1,
        )
        x_0 = np.random.randn(18,4,17,5).astype(np.float32)
        y_0 = DepthToSpace_compute(x_0, 1)
        expect(node, inputs=[x_0], outputs=[y_0], name='test_depthToSpace_0')

        x_0 = np.random.randn(1,1,9,8).astype(np.float32)
        y_0 = DepthToSpace_compute(x_0, 1)
        expect(node, inputs=[x_0], outputs=[y_0], name='test_depthToSpace_1')

        x_0 = np.random.randn(15,2,24,20).astype(np.float32)
        y_0 = DepthToSpace_compute(x_0, 1)
        expect(node, inputs=[x_0], outputs=[y_0], name='test_depthToSpace_2')
        ...
        ...
\end{lstlisting}
\end{table*}
\section{Unit Test Generator}
The unit test generator is a compiler to generate tests in Python. In this section, we focus on explaining: 
\begin{itemize}
\item How to generate a limited number of tests while acquiring a large test coverage;
\item How Sionnx components are organized to generate the test.
\end{itemize}

\subsection{Test Coverage}
It is often unrealistic to generate all valid test cases for one operator. Our goal is trying to generate a limited number of test cases while achieving large test coverage.%which are evenly distributed such that more diverse behaviors of the operator are likely to be tested.

To reach this goal, we propose a three-phase randomization algorithm called TDBc-gen for the test \underline{gen}eration. Our solution is based on the insight that the most important points to differ the code behaviors often reside on the data \underline{T}ype, data \underline{D}imension and \underline{B}oundary conditions of data dimensions and data value. If we further examine the example mentioned in the third challenge in Section~\ref{sec:intro}, the important point is essentially a boundary condition of data dimensions. In this way, if we ensure a full coverage on these three aspects, an overall large test coverage will be more likely to be yielded. In this section, we'll explain TDBc-gen in the following three phases.

\subsubsection{Coverage on Type Combination}
Suppose there are N inputs to be randomized in a test. N[i] represents the i'th input, T[i] represents the allowed type list of i'th input and D[i] represents the number of dimensions of i'th input. How to generate a given number of tests while covering all the type combinations? Our solution is to first calculate total number of type combinations by the equation below:
\begin{equation}
  num\_combinations = \prod_{i=0}^{n-1} len(T[i]) 
\end{equation}

and then randomly generate $\frac{count}{num\_combinations}$ test cases for each type combination.

\subsubsection{Coverage on Data Dimension}
With the number of test cases for each type combination determined, the next step is to randomly generate data such that different number of dimensions will be tested. To avoid ambiguity, we use data\_dim to represent the number of data dimensions. For every generated data, we keep tracking the occurrences of its data\_dim (value must be in range [min\_dim, max\_dim]) during entire randomization. Once the number of occurrences reaches $\frac{count}{max\_dim - min\_dim + 1}$, we mark the current data\_dim as visited and skip it in the rest of randomization. In this way, the generated tests cover all the allowed data dimensions with consistent distribution.

\subsubsection{Coverage on Boundary Conditions}
As pointed out, boundary conditions are very likely to trigger diverse behaviors and are important to be tested. To ensure the selection of boundary conditions, we force TDBc-gen to generate tests to stress on the boundaries of data dimension and data value. 

\subsection{Automatic Test Generation}
Sionnx provides two test profiles for random test generation: "smoke" and "full". The former focuses on generating the operands while keeping the attributes fixed. The latter focuses on generating both the operands and the attributes. 

For the "smoke" profile, we randomly select an instance for each attribute and then split the requested number of test cases (200 by default) through the three-phase TDBc-gen algorithm for the operands. For the "full" profile, we first apply  TDBc-gen for attribute randomization and then follow the same process as the "smoke" profile for the operands.

Below shows two sample commands to generate ONNX tests with Sionnx:

\begin{lstlisting}[language=bash, frame=none]
$ llvm-tblgen -gen-onnx-smoke-tests $osl_file -I $alg_path -o $output_path
$ llvm-tblgen -gen-onnx-tests $osl_file -I $alg_path -o $output_path
\end{lstlisting}
Here, \$osl\_file is the OSL specification file. \$alg\_path is the path of the .algorithm file and  \$output\_path specifies the output folder path for the generated tests. "-gen-onnx-smoke-tests" and "-gen-onnx-tests", as the names indicate, are the options to enable "smoke" and "full" modes respectively.

Listing~\ref{lst:depth-to-space-tests} presents part of tests generated for DepthToSpace by Sionnx. The class definition and the function export() follow the syntax of ONNX test in order to be recognized by the ONNX testing system. The reference function DepthToSpace\_compute (Line 12-16) is copied directly from the DepthToSpace.algorithm file. From line 18-23, it creates an ONNX node with the attribute "blocksize" being randomized to 1, which was done during the compile time in Sionnx. Line 24 shows the randomization of the input x\_0. Line 26 does correctness verification using reference result generated from Line 25. In order to reuse the same computation node among different tests, Sionnx groups all tests into one file. In this example, only three tests are shown.

\section{Conclusion}
ONNX conformance testing is very important to fulfill the ONNX standard. The current testing system relies on handwritten test cases, which is impractical and inefficient. This paper presents the first automatic framework to generate conformance tests for ONNX operators. It proposes an operator specification language to describe the characteristics of the operator. The adoption of other established frameworks in correctness verification helps reduce the burden of writing reference algorithm from scratch. Our randomization strategy is lightweight and able to generate tests with full coverage for data types, data dimensions and boundary conditions, yielding a large test coverage for the implementations of operators.

\bibliographystyle{ACM-Reference-Format}
\bibliography{all} 

%Text of appendix \ldots
\end{document}